\documentclass[pra,showpacs]{revtex4}

\usepackage{dcolumn}
\usepackage{amsmath}
\usepackage{epsfig}
\usepackage{float}

\makeatletter
\def\btt#1{\texttt{\@backslashchar#1}}%
\DeclareRobustCommand\bblash{\btt{\@backslashchar}}%
\makeatother

\begin{document}

\preprint{HEP/123-qed}

\title[Short Title]{Quantum Markov Channels for Qubits}

\author{Sonja Daffer,$^1$ Krzysztof W$\acute{\mbox{o}}$dkiewicz,$^{1,2}$}

\author{John K. McIver$^1$}%
\affiliation{%
    $^1$Department of Physics and Astronomy,
    University of New Mexico,
    800 Yale Blvd. NE,
    Albuquerque, NM 87131   USA  \\
    $^2$Instytut Fizyki Teoretycznej,
    Uniwersytet Warszawski, Ho$\dot{z}$a 69,
    Warszawa 00-681, Poland
    }%


\date{\today}      
\begin{abstract}
\vspace{.1in} \noindent We examine stochastic maps in the context
of quantum optics. Making use of the master equation, the damping
basis, and the Bloch picture we calculate a non-unital, completely
positive, trace-preserving map with unequal damping eigenvalues.
This results in what we call the squeezed vacuum channel.  A
geometrical picture of the effect of stochastic noise on the set
of pure state qubit density operators is provided.  Finally, we
study the capacity of the squeezed vacuum channel to transmit quantum
information and to distribute EPR states.
\\
\\
\end{abstract}


\maketitle

\section{Introduction}
One of the aims of quantum information theory is to achieve the
storage or transmission of information encoded in quantum states
in a fast and reliable way~\cite{bennett1998}. It is unrealistic to
consider a physical system, in which information is stored, as
being isolated. It is well known that when the system of interest
interacts with its environment, irreversible decoherence occurs,
which is, in most cases, both undesirable and
unavoidable~\cite{zurek1991}. This interaction causes pure states
to become mixed states. This process describes the influence of
noise on quantum states which results in information processing
errors.

The question of how to reliably transmit information began with
communication systems.  Shannon's noisy channel coding theorem is
the fundamental theorem of information theory~\cite{shannon1948,cover1991}.
It states that information can be transmitted with arbitrarily
good reliability over a noisy channel provided the transmission
rate is less than the channel capacity, and that a code exists
which achieves this. There has been recent interest in studying
quantum channels for sending quantum information and defining
quantum channel
capacities~\cite{holevo1998,lloyd1997,hausladen1996,bennett1997,
adami1997,barnum1998,shor2002}.

As in classical information theory, a quantum channel capacity is
characterized by the type of noise present in the channel.  There
exists a set of input states or alphabet which the sender
transmits through the channel.  The noise in the channel generally
degrades the states.  The receiver tries to recover the message
which was sent from the output states.  This process of induced
errors may be described by a system interacting with a reservoir.
For a classical communication channel, the channel is completely
characterized by its transition probability matrix which
determines the errors which can occur.  In constrast, a quantum
channel is characterized by a completely-positive,
trace-preserving or stochastic map which takes the input state to
an output state.  This characterizes the type of noise present in
the channel.

In this paper, we use a special basis of left and right damping
eigenoperators for a Lindblad superoperator to calculate explicitly
the image of a stochastic
map for a wide class of Markov quantum channels.
We use this method to derive a noisy quantum channel for qubits which we
call the squeezed vacuum channel.  This channel is non-unital with
unequal damping eigenvalues, which makes it different from
previously introduced channels~\cite{nielsen2000}.  We use this channel to give
geometrical insight into the Holevo channel capacity.

We begin this paper by defining a noisy quantum channel in Section
II in terms of stochastic maps and the Kraus decomposition.  A special
case of the Markov channel is discussed.
In Section III, a general Lindblad equation for a
finite-dimensional  Hilbert space is introduced.  In Section IV,
the damping basis is introduced as an alternative way to
calculate explicitly the stochastic map without using a Kraus
decompostion.  Stochastic maps in the
context of quantum optics described by a set of Bloch equations
are discussed in Section V.  Section VI reviews some known quantum
channels and presents some more general types of channels.
The stochastic map which defines the squeezed vacuum channel is explicitly
calculated
in Section VII and the restrictions imposed by the condition
of complete positivity are presented.  These results are used to determine a
Kraus decomposition explicitly.  The geometrical picture of
the channel is given in Section VIII.  Finally, Section IX
deals with the channel's ability to send
classically encoded quantum states and its ability to send the
resource of entanglement.


\section{Noisy Quantum Channels}
\label{sec:level1}
\subsection{Stochastic Maps}
The concept of a noisy quantum channel arose from the field of
quantum communication. Information is encoded in quantum states
and transmitted across some channel where the receiver decodes the
information to retreive the original message.  The ability to send
messages reliably depends on the noise present in the channel.  The
effect of the noise is to take an initial quantum state and transform
it to another quantum state.  The noisy quantum channel is then defined
by a map
\begin{equation}
    \Phi:\rho\rightarrow\Phi(\rho)
\end{equation}
which takes a quantum state
described by a density operator $\rho$ into a quantum state described
by a density operator $\Phi(\rho)$.  There are certain restrictions on
the class of maps which generate legitimate density operators.  We require
that $Tr[\Phi(\rho)]=1$ so that unit trace of the density operator is conserved for all
time. In addition,
the image of the map, $\Phi(\rho)$, must be a positive operator.  A map which
takes positive operators into positive operators is called a positive map.
But if one considers the noise to come from a larger Hilbert space of a
reservoir, then the stronger condition of complete positivity is required
for the process to be physical~\cite{stinespring1955,choi1972}.  Therefore,
we restrict our attention to
completely positive, trace-preserving maps which are called stochastic maps.

\subsection{The Kraus decomposition}
Noise in the channel may be considered as a reservoir to which the quantum
state of interest is coupled.  The state and the reservoir interact unitarily
for some time and they become correlated.  If we are now only interested in the
system, we trace over the environment degrees of freedom.  One may think of
the reservoir as extracting information from the system as it will typically
map pure states into mixed states.  This noise process can be described by a
quantum operation involving only operators on the system of interest.  This is
called a Kraus decomposition and has the form
\begin{equation}
  \Phi(\rho) = \sum_k A_{k}^{\dagger} \rho A_k
  \label{eq:kraus}
\end{equation}
where the condition
\begin{equation}
    \sum_k A_k A^\dagger_k=I
    \label{eq:identity}
\end{equation}
ensures that unit trace is
preserved for all time~\cite{kraus1983}.
If an operation has a Kraus decomposition, then it is completely positive.
The converse is also true so that all stochastic maps have a Kraus decomposition.

\subsection{The Lindblad form}

The formalism we have outlined so far is general.  For a important type of noise,
Markov noise, we have a special class of completely positive maps.
We call a Markov quantum channel one in which the noise in the channel arises from a coupling
of the system with a reservoir under the Markov and Born approximations.  This is a
commonly used approximation in quantum optics and leads to the well-known Lindbladian
form of a master equation.
For this type of channel, one can always write a stochastic map as
\begin{equation}
    \Phi(\rho)=e^{{\cal L}t} \rho(0).
\end{equation}
The equation describes the evolution of a system coupled to a reservoir in terms
of the system of interest alone.  All Lindblad superoperators are stochastic maps
and have a Kraus decomposition.  The converse is not true in general.

In this paper, we will derive an equivalent equation for $\Phi$ of the
form
\begin{equation}
    \Phi(\rho)=\sum_i Tr \lbrace {\mbox R_i} \rho(0) \rbrace \Lambda_i {\mbox
  L_i}
\end{equation}
to obtain the image of the stochastic map for Markov noise.  We use a special basis
of left, $L_i$, and right, $R_i$, eigenoperators which allow for an explicit calculation of the stochastic
map and the Kraus operators.  This method works for any Markov channel.

\section{The General Lindblad Equation}
\label{sec:level1}

The Schr$\ddot{{\mbox o}}$dinger evolution of a system coupled to a
reservoir
can be described in terms of a master equation of the form
\begin{equation}    \label{eq:master}
  \dot{\rho}= {\cal L} \rho
  ={\cal L}_\textsf{C} \rho+{\cal L}_\textsf{D} \rho
\end{equation}
where $\rho$ is the density operator of the system alone, obtained
by tracing over the reservoir degrees of freedom. The first term
describes the coherent or unitary evolution and is simply given by
the commutator
\begin{equation}    \label{eq:commutator}
   {\cal L}_\textsf{c} \rho = -\frac{i}{\hbar} \lbrack H , \rho \rbrack
\end{equation}
where H is the Hamiltonian of the undamped system.

The general
form of the non-unitary part which describes the dissipation
of the density operator is
\begin{equation}
  {\cal L}_\textsf{D} \rho =  \frac{1}{2} \sum_{i,j}^{N^2-1}
  c_{i,j}  \lbrace [F_i,\rho F^\dagger_j] +[F_i \rho,
  F^\dagger_j] \rbrace
  \label{eq:generallindblad}
\end{equation}
valid for a finite N-dimensional Hilbert
space. The $\lbrace F_i \rbrace$ are system
operators which satisfy the conditions ${\mbox {Tr}(F_i^\dagger
F_j)}=\delta_{i,j}$ and ${\mbox {Tr}(F_i)}=0$. The set of complex
elements $\{c_{ij}\}$ form a positive matrix.

It has been proven~\cite{kossakowski1976} that a linear operator on a finite
N-dimensional Hilbert space ${\cal L}:M(N)\rightarrow M(N)$ is the
generator of a completely positive dynamical semigroup in the
Schr\"{o}dinger picture if and only if it can be written in the
form of ${\cal L}: \rho \rightarrow {\cal L} \rho$ where ${\cal L}
\rho$ takes the form of Eqs. (\ref{eq:commutator}) and
(\ref{eq:generallindblad}).  We will
call the generator of the semigroup which governs the dissipation
the Lindbladian and denote it by ${\cal L}_\textsf{D}$. Eq.
(\ref{eq:generallindblad}) may be recognized as the master
equation describing irreversible evolution of an open quantum
system under the Markov and Born approximations.
This Lindblad equation is widely used in many branches of
statistical mechanics and quantum optics. This form, the Lindblad
form, has been shown to guarantee positivity and
trace-preservation of the density
operator~\cite{alicki1987,lindblad1983}.

\section{The Damping Basis}
There are many methods for solving master equations, the use of
Fokker-Planck equations built on methods in stochastic processes,
and Monte-Carlo wave functions, to name a couple~\cite{walls1994,gardiner1991}.
In this paper,
we will make use of the damping basis in order to solve a
master equation which has the form of Eq. (\ref{eq:master})
containing both the coherent and damping dynamics.  This amounts
to solving an eigenvalue equation. In some cases, this problem can
seem formidable and finding a damping basis first is useful.  To
solve the master equation in this fashion involves first solving
the eigenvalue equation
\begin{equation}
  {\cal L}_\textsf{D} \rho = \lambda \rho      \label{eq:eigenvalue}
\end{equation}
for the non-unitary part of the density operator evolution
describing an open system.  This provides one with a complete,
orthogonal basis with which to expand the density operator at any
time.  Such a basis is called the damping basis~\cite{briegel993}.
This basis is obtained by finding the eigenoperators of the
eigenvalue equation.  Likewise, the dual eigenoperators are found by
solving the dual eigenvalue equation.  The original basis and the
dual basis are orthogonal.

If the eigenoperators of Eq. (\ref{eq:eigenvalue}) are ${\mbox
R_i}$ with corresponding eigenvalues $\lambda_i$ then once the
initial state is known
\begin{equation}
  \rho(0)= \sum_{i} Tr \lbrace {\mbox L_i} \rho(0) \rbrace
  {\mbox R_i},
\end{equation}
the state of the system at any later time can be found through
\begin{eqnarray}    \label{eq:rhooft}
  \rho(t)&=&e^{{\cal L}t} \rho(0)    \nonumber  \\
  &=&\sum_i Tr \lbrace {\mbox L_i} \rho(0) \rbrace \Lambda_i {\mbox
  R_i}  \\
  &=&\sum_i Tr \lbrace {\mbox R_i} \rho(0) \rbrace \Lambda_i {\mbox
  L_i}   \nonumber
\end{eqnarray}
where $\Lambda_i = e^{\lambda_i t}$ are the damping eigenvalues
and ${\mbox L_i}$ is the state dual to ${\mbox R_i}$. These are
called the left and right eigenoperators, respectively, and
satisfy the following duality relation:
\begin{equation}
  Tr \lbrace {\mbox L_i}  {\mbox R_j} \rbrace = \delta_{ij}.
  \label{eq:duality}
\end{equation}
It is easy to show that {L} and {R} have the same eigenvalues. The
left eigenoperators satisfy the eigenvalue equation
\begin{equation}  \label{eq:lefteigen}
  {\mbox L}{\cal L}_\textsf{D} =\lambda {\mbox L}
\end{equation}
while the right eigenoperators satisfy
\begin{equation}  \label{eq:righteigen}
  {\cal L}_\textsf{D} {\mbox R} = {\mbox R} \lambda.
\end{equation}
This method is a simple way of finding the density operator for a
given ${\cal L}$ for all times.  The solution of the left and
right eigenvalue equations yields a set of eigenvalues and
eigensolutions:
$\lbrace \lambda, {\mbox L}, {\mbox R} \rbrace.$
Once the damping basis is obtained, it can be used to expand the
density operator.  Then the density operator in this basis can be
substituted back into the full Liouville Eq. (\ref{eq:master}).
By doing this, one obtains a set of coupled differential equations
for the coefficients of the density operator in the damping basis.
Solution of this set of coupled differential equations yields the
solution to the total Liouville dynamics.  The important point is
that once all eigenvalues and all left and right eigenoperators of
the superoperator are found, the master equation can be solved and
all system observables can be computed.

\section{Bloch Stochastic Map}

When studying two-level systems there is the added advantage of a
geometrical picture offered by the vector model of the density
matrix.  For instance, decoherence of a two-level atom is
described by the dynamics of a Bloch vector with three components,
$\vec{b}=(u,v,w)$, inside a unit three-sphere, governed by a set
of Bloch equations~\cite{allen1975}.  These constitute a set of
differential equations, one for each component of the Bloch
vector, of the form:
\begin{eqnarray}  \label{eq:bloch}
 \dot{u}&=&-\frac{1}{T_u} u-\Delta v               \nonumber\\
 \dot{v}&=&-\frac{1}{T_v} v+\Delta u + \Omega w      \\
 \dot{w}&=&-\frac{1}{T_w} (w-w_{eq})- \Omega v,      \nonumber
\end{eqnarray}
where $\Omega$ is the Rabi frequency, the constants $T_u$ and
$T_v$ are decay rates of the atomic dipole, and $T_w$ is the decay
rate of the atomic inversion into an equilibrium state $w_{eq}$.
One typically finds that the phenomenological decay rates in the
Bloch equations appear as
\begin{equation}  
   \frac{1}{T_u} = \frac{1}{T_2}, \hspace{.3in}
   \frac{1}{T_v} = \frac{1}{T_2}, \hspace{.3in}
   \frac{1}{T_w} = \frac{1}{T_1}, \hspace{.3in}
\end{equation}
so that the parts of the atomic dipole which are in phase and out
of phase with the driving field are affected in the same way by
the damping. This description is in terms of a two-level atom
coupled to an external field as well as a reservoir.  The coupling
to the external field causes the Bloch vector to rotate.  The
coupling to the reservoir, which might be a continuum of vacuum
field modes, causes the Bloch vector to decrease in magnitude. The
combination of these two behaviors leads to a spiraling in of the
Bloch vector.  Although we describe these dynamics in terms of the
two-level atom, the Bloch picture can describe any two-level
system. Here we shall consider the more general form of Eq.
(\ref{eq:bloch}) where all three damping constants may be unequal.
As we shall see, this describes the physical situation where the
two-level atom is coupled to a squeezed vacuum reservoir rather
than a regular vacuum field.  The damping parameters lead to
decoherence of the system of interest.

The decoherence is caused by the presence of noise and may be
viewed as a stochastic map acting on the Bloch vector in the
form of a mapping~\cite{wodkiewicz2001}:
\begin{equation}
   \Phi : {\vec b} \rightarrow {\vec b'}.
   \label{eq:phionb}
\end{equation}
Because there is a correspondence between the Bloch vector
$\vec{b}$ and the density operator $\rho$, we see that the
stochastic map is a superoperator which maps density operators
into density operators:
\begin{equation}
   \Phi : \rho \rightarrow \rho'.
   \label{eq:phionrho}
\end{equation}
We can expand the density
operator in the Pauli basis, $\lbrace I,\sigma_x,\sigma_y,\sigma_z
\rbrace$, and consider how the components of $\rho$ transform
under the map. This latter transformation is characterized by a
4x4 matrix representation of $\Phi$. It has been found that the
general form of any stochastic map on the set of complex 2x2
matrices, may be represented by such a 4x4 matrix containing 12
parameters~\cite{ruskai2002}:
\begin{equation}   \label{eq:tmatrix}
  {\cal T}=
  \left(
  \begin{array}{cccc}
    1        &       0    &    0     &     0    \\
    t_{10}   &    t_{11}  &  t_{12}  &  t_{13}  \\
    t_{20}   &    t_{21}  &  t_{22}  &  t_{23}  \\
    t_{30}   &    t_{31}  &  t_{32}  &  t_{33}
  \end{array}
  \right).
\end{equation}
The 3x3 block of the matrix ${\cal T}$ can be diagonalized using
two rotations.  This amounts to a change of basis.  Without loss
of generality, we can consider the matrix
\begin{equation}    \label{eq:tdmatrix}
  {\cal T_{\cal D}}=
  \left(
  \begin{array}{cccc}
    1        &       0      &       0      &        0     \\
    t_{10}   &  \Lambda_1   &       0      &        0     \\
    t_{20}   &       0      &   \Lambda_2  &        0     \\
    t_{30}   &       0      &       0      &    \Lambda_3
  \end{array}
  \right)
\end{equation}
which uniquely determines the map.  To preserve hermiticity,
${\cal T}$ must be real.  The first row must be $\lbrace 1,0,0,0
\rbrace$ to preserve the trace of the density operator.  We call
the 3x3 part of the matrix ${\cal T_{\cal D}}$, consisting of the
damping eigenvalues $\Lambda_i$ which are contractions, the
damping matrix $\bf \Lambda$.  Explicitly,
\begin{equation}
  {\bf \Lambda} =
  \left(
  \begin{array}{ccc}
     \Lambda_1       &    0           &     0         \\
         0           &  \Lambda_2     &     0         \\
         0           &    0           &   \Lambda_3
  \end{array}
  \right).          \label{eq:dampingmatrix}
\end{equation}
In terms of the Bloch vector, a general stochastic map may be written in
the form
\begin{equation}
  \Phi : \vec{b} \rightarrow \vec{b'} = {\bf \Lambda} \vec{b} + \vec{b_o}
\end{equation}
where $\vec{b_o}= \left( t_{10}, t_{20}, t_{30} \right)$ is a
translation.  The overall operation consists of a damping part and
translations.  Due to the presence of translations, the
transformation is affine.

To see the properties required by the stochastic map in terms of
the Bloch vector consider the matrix representation of the Bloch
vector as an expansion in terms of the Pauli matrices:
\begin{equation}
  B=\vec{b} \cdot \vec{\sigma} =
  \left(
  \begin{array}{cc}
    w         &        u-iv  \\
    u+iv      &        -w
  \end{array}
  \right).
\end{equation}
In the absence of noise, the Bloch vector remains on the Bloch sphere so
that
\begin{equation}
   \mbox{det B} = - (u^2 + v^2 +w^2)
\end{equation}
has magnitude unity.
The presence of stochastic noise transforms the matrix $B$ according to
\begin{equation}
   \Phi:B \rightarrow B'.
\end{equation}
To guarantee that the map $\Phi$ transforms the density operator
into another density operator, the Bloch vector can only be
transformed into a vector contained in the interior of the Bloch sphere,
or the Bloch ball.  Equivalently, we require
\begin{equation}
   |\mbox{det B}'| \leq |\mbox{det B}|
\end{equation}
so that the qubit density operator
\begin{equation}
   \rho=\frac{1}{2} \left( I+\vec{b} \cdot \vec{\sigma} \right) =
   \frac{1}{2}
   \left( I + B \right)
\end{equation}
under the map becomes
\begin{equation}
   \Phi(\rho)  : \rho \rightarrow \Phi (\rho) =\frac{1}{2} \left( I + B'
   \right).
\end{equation}
This is only possible if the eigenvalues $\Lambda_i$ are
contractions. In the following sections we provide an explicit
construction of the stochastic map $\Phi$ from the Bloch Eqs.
(\ref{eq:bloch}).

\section{Two-level atom in a Squeezed Vacuum}

The master equation (\ref{eq:master}) yields a plethora of
possible completely positive dynamical maps.  In the remainder of
this paper, we wish to examine a particular form of Eq.
(\ref{eq:master}) which retains a relation to the Bloch Eqs.
(\ref{eq:bloch}) and leads to the case where the three components
of the Bloch vector have different decay rates and where the Bloch
vector is shifted from the origin of the Bloch sphere.  As will be
seen, this leads to a contraction of the set of states which lie
on the Bloch sphere surface.  Some states in the set become very
mixed, while some remain almost pure.

We will consider a special case of the Lindbladian in
Eq.(\ref{eq:generallindblad}) for a two-dimensional Hilbert space
by choosing the following set of system operators $\{F_i\}$:
\begin{equation}
\label{eq:Fs}
  F_1 = \sigma,  \hspace{.1in}
  F_2 = \sigma^\dagger,  \hspace{.1in}
  F_3 = \frac{\sigma_3}{\sqrt{2}}
\end{equation}
where $\sigma$ and $\sigma^\dagger$ are the qubit lowering and raising
operators and $\sigma_3$ is the z-component
Pauli spin operator.
If, along with this set of system operators, we choose the matrix
elements ${c_{ij}}$ such that
\begin{equation}
\textbf{c}= \left(
\begin{array}{ccc}
\frac{1}{2T_1}(1-w_{eq}) &   -\frac{1}{T_3}    & 0       \\
 -\frac{1}{T_3}          & \frac{1}{2T_1}(1+w_{eq})   & 0  \\
0                        & 0                     &
\frac{1}{T_2}-\frac{1}{2T_1}
\end{array}
\right)          \label{eq:cmatrix}
\end{equation}
%
the resulting Lindbladian is of the form
\begin{eqnarray}
\label{eq:lindblad}
  {\cal L}_\textsf{D}\rho = & - & \frac{1}{4T_1}(1-w_{eq})
  [\sigma^\dagger\sigma\rho+\rho\sigma^\dagger\sigma-2\sigma\rho
  \sigma^\dagger]
  \nonumber\\
  &-&  \frac{1}{4T_1}(1+w_{eq})[\sigma\sigma^\dagger\rho+
  \rho\sigma\sigma^\dagger-2\sigma^\dagger\rho\sigma]
  \nonumber\\
  &-& \left( \frac{1}{2T_2}-\frac{1}{4T_1} \right)[\rho-\sigma_3\rho\sigma_3]
  \\
  &-&\frac{1}{T_3}[\sigma^\dagger\rho\sigma^\dagger
  +\sigma \rho\sigma ].  \nonumber
\end{eqnarray}
This part of the master equation is known to describe the
dissipative evolution of a two-level atom coupled to a
bath~\cite{walls1994}. The raising and lowering operators describe
transitions between the ground and excited states and the
${c_{ij}}$ describe the losses caused by the reservoir and depend
on phenomenological decay constants.

The coherent part of the dynamics, ${\cal L}_\textsf{C}$, will be
described by the Hamiltonian
\begin{equation}
  H=\frac{\hbar \Omega}{2} (\sigma^\dagger + \sigma)
  \label{eq:hamiltonian}
\end{equation}
where $\Omega$ is the Rabi frequency of oscillation between the
ground and excited states.
The full dynamics describe a two-level atom driven by a laser
field subjected to irreversible decoherence by its environment.
This corresponds to a linear mapping from a two-dimensional
Hilbert space into a two-dimensional Hilbert space.  More
generally, this Liouvillian generates a completely positive
dynamical map of a generic two-level system or qubit.

The Bloch equations for a two-level system described by this
Lindblad operator (\ref{eq:lindblad}) are given by Eqs.
(\ref{eq:bloch}) where
\begin{eqnarray}      \label{eq:damping}
  \frac{1}{T_u}&=&\frac{1}{T_2}+\frac{1}{T_3}  \nonumber  \\
  \frac{1}{T_v}&=&\frac{1}{T_2}-\frac{1}{T_3}   \\
  \frac{1}{T_w}&=&\frac{1}{T_1}.               \nonumber
\end{eqnarray}
The presence of the parameter $T_3$ is the source of
the damping asymmetry between the $u$ and $v$ components of the
Bloch vector.
Eq. (\ref{eq:lindblad}) describes many well-known physical
processes in quantum optics.  We wish to now briefly discuss the
physics behind this Lindblad form of the master equation.
\subsection{The Amplitude Damping Channel}
The amplitude damping channel is identical to the case of
spontaneous emission for a two-level atom.  This corresponds
to the following choice for the parameters:
\begin{equation}
 \frac{1}{T_1} = A,                 \hspace{.2in}
 \frac{1}{T_2} = \frac{A}{2},       \hspace{.2in}
 \frac{1}{T_3} = 0,                 \hspace{.2in}
 w_{eq} = -1.   \label{eq:sponemission}
 \end{equation}
The parameter A is the Einstein coefficient of spontanteous
emission that depends on the density of vacuum field modes and how
strongly the atom couples to the modes.   This describes the
exponential decay of an atom, from excited to ground state, due to
vacuum fluctuations.  The equilibrium state is $w_{eq} = -1$
indicating that, given enough time, the atom will be in the ground
state.

\subsection{The Depolarizing Channel}
Another type of noisy quantum channel is the depolarizing channel.
This channel is identical to the quantum optical model for pure phase decay.
This channel has parameters
\begin{equation}   \label{eq:phasedecay}
 \frac{1}{T_1} = 0,                          \hspace{.2in}
 \frac{1}{T_2} = \Gamma,       \hspace{.2in}
 \frac{1}{T_3} = 0,                          \hspace{.2in}
 w_{eq} = 0.
\end{equation}
This describes the process of phase randomization of the atomic
dipole caused by atomic collisions.  This leads to equal damping
for the $u$ and $v$ components of the Bloch vector with
contractions which depend on the parameter, $\Gamma$, due to
collisions.

\subsection{The Thermal Field Channel}
In the case of spontaneous emission, the atom is coupled to a vacuum
reservoir.  But one can consider an atom interacting with a thermal field,
so that now the field has a non-zero photon number.
The reservoir may be considered as a large number of harmonic
oscillators such as modes of the free electromagnetic field or a heat bath
in equilibrium.  In this case, one finds the decay constants are related
to the photon number in the following way:
\begin{eqnarray}
   \frac{1}{T_1} &=& 2A \left( N+\frac{1}{2} \right),  \hspace{.1in}
   \frac{1}{T_2} = A \left( N+\frac{1}{2} \right), \\
   \frac{1}{T_3} &=& 0,    \hspace{.1in}
   w_{eq} = -\frac{1}{2N+1}.  \nonumber
\end{eqnarray}
Note that setting $N=0$ reduces to the case of spontaneous
emission.  In this thermal field case, the value of $w_{eq}$
indicates that the atomic inversion approaches a steady state
which is the ground state in the limit of zero photon number but
as N becomes large, it approaches zero.  Therefore, the
equilibrium state for the inversion is bounded: $-1<w_{eq}<0.$ One
can see from Eqs. (\ref{eq:damping}) that this leads to equal
damping for the $u$ and $v$ components of the Bloch vector.

\subsection{The Squeezed Vacuum Channel}
A more general case occurs when all parameters are non-zero and
explicitly are:
\begin{eqnarray}
 \frac{1}{T_1} &=& 2A(N+\frac{1}{2}),   \hspace{.1in}
 \frac{1}{T_2} = A(N+\frac{1}{2}), \\
 \frac{1}{T_3} &=& A M,              \hspace{.1in}
 w_{eq} = -\frac{1}{2N+1}.  \nonumber
\end{eqnarray}
This describes an atom in a squeezed vacuum where N is the photon
number and M is the squeezing parameter.  The parameter N is related
to the two-time correlation function for the noise operators of the
reservoir $\langle a^\dagger(t) a(t') \rangle = N \delta (t-t')$ where
$a(t)$ is the field amplitude for a reservoir mode.  The squeezing
parameter, M, arises from the two-time correlation function involving
the square of the field amplitudes $\langle a (t) a(t') \rangle = M^\star \delta (t-t')$.
These are the familiar relations obeyed by squeezed white noise which leads to
squeezing of a vacuum reservoir~\cite{gardiner1991}.
A squeezed vacuum has
fluctuations in one quadrature smaller than allowed by the
uncertainty principle at the expense of larger fluctuations in the
other quadrature.  This type of reservoir leads to two dipole
decay constants in the Bloch equations, the one in the squeezed
quadrature being correspondingly smaller than that for the
stretched quadrature.

The steady state for the inversion is the same here as in the
previous case and depends only on the photon number.  When the
field is not squeezed ($M=0$), the master equation reduces to the
previous case of an atom in a thermal field. The important
difference now is that the squeezing parameter has introduced a
new parameter, $T_3,$ which leads to unequal damping for the $u$
and $v$ components of the Bloch vector.  In what follows, we deal
with the most general case, corresponding to the Lindbladian
(\ref{eq:lindblad}), of a qubit coupled to a squeezed vacuum
reservoir.  This defines a new noisy quantum channel which we call
the squeezed vacuum channel (SVC) having different properties than
the depolarizing and amplitude damping channels.
\section{The Squeezed Vacuum Channel}
\subsection{The Image of the Map}
So far, we have introduced a general set of Bloch Eqs.
(\ref{eq:bloch}) which correspond to the Liouvillian with Eq.
(\ref{eq:lindblad}) in addition to the coherent dynamics described
by $H$ in Eq. (\ref{eq:hamiltonian}). We are now in a position to
solve this master equation. We proceed with the method described
in section IV. This tells us how the noise affects all states of
the two-level system. Using the damping basis, the solution to the
Liouville Eq. (\ref{eq:eigenvalue}) with Lindblad operator of the
form of Eq. (\ref{eq:lindblad}) follows. For a two-level system in
a squeezed vacuum reservoir, the left eigenoperators are
\begin{eqnarray}
  L_0&=&\frac{1}{\sqrt{2}} I ,   \hspace{.5in}
  L_1=\frac{1}{\sqrt{2}}(\sigma^\dagger+\sigma), \\
  L_2&=&\frac{1}{\sqrt{2}}(\sigma^\dagger-\sigma),   \hspace{.1in}
  L_3=\frac{1}{\sqrt{2}}(-w_{eq}I+\sigma_3) \nonumber
\end{eqnarray}
found from solving Eq. (\ref{eq:lefteigen}) while the right
eigenoperators are
\begin{eqnarray}
  R_0&=&\frac{1}{\sqrt{2}} (I + w_{eq} \sigma_3),   \hspace{.1in}
  R_1=\frac{1}{\sqrt{2}}(\sigma^\dagger+\sigma),  \\
  R_2&=&\frac{1}{\sqrt{2}}(\sigma-\sigma^\dagger),  \hspace{.3in}
  R_3=\frac{1}{\sqrt{2}} \sigma_3  \nonumber \label{eq:rdampingbasis}
\end{eqnarray}
found by solving Eq. (\ref{eq:righteigen}). They correspond to the
following four eigenvalues:
\begin{eqnarray}
   \lambda_0&=&0, \hspace{.3in}
   \lambda_1= - \left(\frac{1}{T_2}+\frac{1}{T_3}\right), \\
   \lambda_2&=& - \left(\frac{1}{T_2}-\frac{1}{T_3}\right) ,
   \hspace{.1in}
   \lambda_3= - \frac{1}{T_1} .    \nonumber  \label{eq:evalues}
\end{eqnarray}
Of the four eigenvalues, three of them are precisely the diagonal
elements of the damping matrix (\ref{eq:dampingmatrix}) through
the equation $\Lambda = e^{\lambda t}$. We call the set
$\{\Lambda\}$ the damping eigenvalues and the set $\{\lambda\}$,
the eigenvalues (of the damping basis). This indicates a relation
between the decay constants in the Bloch equations and the damping
basis. The damping eigenvalues of the damping matrix $\Lambda$ contain
the decay constants for the three components of the Bloch vector.
The fourth of the damping eigenvalues is unity, which is a
necessary condition for the density operator to have trace unity.
The density operator may be expanded in any complete basis.
Choosing the right eigenoperators as a basis we can write the
density operator as
\begin{equation}
    \rho=\emph{l}_0 R_0 +\emph{l}_1 R_1+\emph{l}_2 R_2+\emph{l}_3 R_3
\end{equation}
where the coefficients are obtained by projecting on to the left
eigenbasis,
$\emph{l}_i={\mbox {Tr}} \{ L_i \rho \}.$
Next we substitute the expansion on the right eigenoperators into
the equation for the total Liouville operator equation and use the
fact that ${\cal L}_\textsf{D} R_i = \lambda_i R_i.$  After the
substitution of $ \rho$ into the total Liouville equation, one
obtains a set of differential equations for the coefficients
$\emph{l}_i$
in the right eigenbasis and the generator of the Markovian time
evolution has the following matrix representation in the same
basis
\begin{equation}
{\cal L}= \left(
\begin{array}{cccc}
\lambda_0                         & 0             & 0                     & 0  \\
        0                         & \lambda_1     & 0                     & 0  \\
-i w_{eq}\Omega                 & 0             & \lambda_2             &  -i \Omega  \\
        0                         & 0             & -i \Omega            & \lambda_3
\end{array}
\right).         \label{eq:Mdampingbasis}
\end{equation}
The solution to $\dot{\rho}={\cal L} \rho $ is
\begin{equation}
    \rho(t)= e^{{\cal L}t}  \rho(0).
\end{equation}
One can perform a rotation of the right eigenbasis to obtain the
following diagonalized form of the above matrix:
\begin{equation}
{\cal L}= \left(
\begin{array}{cccc}
\lambda_0      & 0              & 0               & 0  \\
0              & \lambda_1      & 0               & 0  \\
0              & 0              & \lambda_{23}+\chi    & 0  \\
0              & 0              & 0               &
\lambda_{23}-\chi
\end{array}
\right)          \label{eq:Mrotateddampingbasis}
\end{equation}
where $\lambda_{23}= \frac{\lambda_2+\lambda_3}{2}$,
$\chi=\frac{1}{2} \sqrt{(\lambda_2-\lambda_3)^2-(2\Omega)^2},$ and
the $\lambda_i$ are given in Eq. (\ref{eq:evalues}).  It follows
that the superoperator $e^{{\cal L}t}$ which maps the density
operator forward in time is
\begin{equation}
e^{{\cal L} t}= \left(
\begin{array}{cccc}
1              & 0              & 0               & 0  \\
0              & \Lambda_1      & 0               & 0  \\
0              & 0              & \Lambda_{23}e^{\chi t}    & 0  \\
0              & 0              & 0               &
\Lambda_{23}e^{-\chi t}
\end{array}
\right)          \label{eq:superoperator}
\end{equation}
The matrices above are represented in two different bases. Eq.
(\ref{eq:Mdampingbasis}) is in the damping basis, while Eq.
(\ref{eq:Mrotateddampingbasis}) is in a rotated damping basis.
Geometrically, one may consider the first case to be dynamics as
viewed from a shifting center of the Bloch sphere.  In this case,
there is one affine shift towards the South pole of the Bloch
sphere. This is the viewpoint in the damping basis. This can be
seen by noting that in Eqs. (\ref{eq:rdampingbasis}) for the
damping basis there is one shift present in $\mbox{R}_0$. From the
right eigenoperators of the damping basis, one finds that they are
almost the same as the Pauli basis. From the viewpoint of the
Pauli basis, the dynamics would take place as seen from the
stationary center of the Bloch sphere.  In the rotated damping
basis, the viewpoint is from a frame which is rotating with the
driving field as well as shifting from the center of the Bloch
sphere and, consequently, the eigenvalues lead to contractions or
pure damping in this diagonal basis.  Although the most general
dynamics contains both coherent and incoherent parts, in certain
situations one part may dominate the dynamics. We will consider
the case where the system is not isolated from its environment but
unaffected by coherent dynamics. In this case, rotations occur on
a time scale much longer than the dissipation so that effectively
$\Omega \rightarrow 0.$  We will see that this case is
advantageous.

With these explicit formulas for the damping basis and the damping
eigenvalues we can calculate the density operator for all times.
This gives us the image of the map, $\Phi(\rho).$ Assuming the
initial density matrix is of the form
\begin{equation}
   \rho=
   \left(
   \begin{array}{cc}
     a & d  \\
     d^\star & c
   \end{array}
   \right)
\end{equation}
we find that the stochastic map generates a new density matrix
\begin{equation}
   \Phi(\rho)=
   \left(
   \begin{array}{cc}
     A & D  \\
     D^\star & C
   \end{array}
   \right)  \label{eq:image}
\end{equation}
in accordance with Eq. (\ref{eq:phionrho}).  This is obtained
using the set the eigenoperators and eigenvalues for the damping
basis. For the stochastic map which characterizes the squeezed
vacuum channel, we have that the elements of $\Phi(\rho)$ are
given by
\begin{eqnarray}
  A\hspace{.05in}&=&\hspace{.05in}\frac{1}{2} (a+c)(1+w_{eq}) +
  \frac{1}{2}
  \Lambda_3(a-c-w_{eq}(a+c)),
  \nonumber\\
  C\hspace{.05in}&=&\hspace{.05in}\frac{1}{2} (a+c)(1-w_{eq}) -
  \frac{1}{2}
  \Lambda_3(a-c-w_{eq}(a+c)),
  \nonumber \\
  D\hspace{.05in}&=&\hspace{.05in}\frac{1}{2} \lbrack
  d^\star(\Lambda_1-\Lambda_2) + d
  (\Lambda_1+\Lambda_2) \rbrack,
  \\
  D^\star&=&\hspace{.05in} \frac{1}{2} \lbrack d (\Lambda_1-\Lambda_2) +
  d^\star
  (\Lambda_1+\Lambda_2) \rbrack.\nonumber
\end{eqnarray}
These elements are in terms of the initial density matrix elements
as well as the damping eigenvalues which contain the parameters
${T_i},$ and $w_{eq}.$  The image gives us the density matrix
after the noise operator $\Phi$ has acted on it.  The channel
capacity of a noisy quantum channel is characterized by the types
of errors which result after the input is transmitted.  The noise
operation defines the channel.

\subsection{Complete Positivity}
The restriction that this map be completely positive is somewhat
stringent.  We have mentioned so far that the operation of $\Phi$
on the Bloch vector must transform the vector into another vector
inside the Bloch sphere.  This is just one rather obvious
condition.  What is not as obvious is that not all states inside
the Bloch sphere are allowable for the system.  In other words,
the Bloch vector can not access all points in the interior of the
Bloch sphere.  This is because of the condition of complete
positivity, related to the existence of a Kraus representation
which we discuss in the following section.

It has been shown~\cite{ruskai2001} that the damping eigenvalues must obey
the four inequalities
\begin{eqnarray}    \label{eq:inequalities}
  \Lambda_1 + \Lambda_2 - \Lambda_3 \leq 1
  \nonumber\\
  \Lambda_1 - \Lambda_2 + \Lambda_3 \leq 1
  \\
  -\Lambda_1 + \Lambda_2 + \Lambda_3 \leq 1
  \nonumber\\
  -\Lambda_1 - \Lambda_2 - \Lambda_3 \leq 1
  \nonumber
\end{eqnarray}
to guarantee complete positivity of the map.  This is a necessary
condition.  These inequalities are the most general case, {\it
i.e.}, they apply to the damping matrix no matter which case is
considered. These inequalities are a consequence of the set of
equations given in the Appendix which involve the damping matrix
elements $\Lambda_i$.  There are five such equations which, taken
with the first equation, are inner products of vectors.  The fact
that the inner product must be positive semi-definite leads to the
inequalities above. For the specific example we present, they
reduce to a more familiar form. Because for spontaneous emission
we have $ 2 T_1 = T_2$, the four inequalities become:
\begin{eqnarray}
  \pm \cosh \left( \frac{t}{T_3} \right) \leq \cosh \left( \frac{t}{T_2}
  \right)  \\
  \pm \sinh \left( \frac{t}{T_3} \right) \leq \sinh \left( \frac{t}{T_2}
  \right)  \nonumber
\end{eqnarray}
which are satisfied if and only if
\begin{equation}
  {\Big |} \frac{1}{T_3} {\Big |} \leq {\Big |} \frac{1}{T_2} {\Big |}.
\end{equation}
In terms of the parameters M and N, this condition for complete positivity
becomes
\begin{equation}
  {\Big |} M {\Big |} \leq    N + \frac{1}{2}
\end{equation}
as expected.  It is well known that M and N are not independent
and the amount of squeezing is limited by the number of photons. A
stricter inequality can be derived directly from the matrix
elements $c_{ij}$ of the Lindbladian form in
Eq.(\ref{eq:generallindblad}).  This must be a non-negative
matrix.  Using the matrix elements of Eq.(\ref{eq:cmatrix}) one
finds that the c-matrix is non-negative if ${\mbox M}^2 \leq
{\mbox {N(N+1)}}$ where N is the photon number and M is the
squeezing parameter.  In the case of equality, we have pure
squeezing.  While the general inequalities of Eq. (\ref{eq:inequalities}) are a necessary
condition for complete positivity, the stricter inequality ${\mbox M}^2 \leq
{\mbox {N(N+1)}}$ for the SVC
are both necessary and sufficient.

It is worth pointing out that
while contractions, without shifts, will always map the Bloch
sphere into the Bloch ball, not all contractions satisfy the
inequalities (\ref{eq:inequalities}).  Those which do not satisfy
them, do not have a Kraus representation and, hence, are not
completely positive maps.  We shall see that the condition of
complete positivity restricts the allowable ellipsoids of the
image.  Because the Lindblad form of the master equation
guarantees complete positivity, by Kraus' theorem there must exist
a decomposition with Kraus operators which acts on the system of
interest alone.  In the next
section, we give an explicit form of the Kraus operators.

\subsection{The Kraus Decomposition}
For the special case of a qubit there can be at most four Kraus
operators.  To see why this is true, the reader is referred to a
lemma where it is shown that the minimum number of Kraus operators
is equal to the rank of a certain matrix~\cite{ruskai2001}. It
follows that the minimum number of Kraus operators needed to
represent the map for the squeezed vacuum channel is four. We do
not show this, but rather, use this fact to find an operator-sum
representation using the minimum number of Kraus operators
possible.  We start by assuming that the four Kraus operators
exist and that they are real, although in general they are
represented by complex matrices. Next we expand all 2x2 matrices
in Eq. (\ref{eq:kraus}) in the Pauli basis. If we let
\begin{eqnarray}   \label{eq:krausoperators}
  A_{k} &=& m_{k0} I +\vec{m_k} \cdot \vec{\sigma}
  \hspace{.05in}=  A_k\left( \vec{m}
  \right)           \\
  A_{k}^{\dagger} &=& m_{k0}^{\star} I +\vec{m_k}^{\star} \cdot
  \vec{\sigma} = A_k^{\dagger} (\vec{m^{\star}})  \nonumber
\end{eqnarray}
then Eq. (\ref{eq:kraus}) becomes
$
  \frac{1}{2} \left( 1 + s_i(a,c,d,d^\star) \sigma_i \right)  =
  \frac{1}{2} \left( 1 + t_i ( \vec{m},\vec{m^\star},a,c,d,d^\star ) \sigma_i
  \right).
$ Equating each coefficient in $s_i$ and $t_i$ leads to a set of
linear equations which underdetermine the sixteen coefficients of
the Kraus operators in Eq. (\ref{eq:krausoperators}) in the Pauli
basis.  Thus, the Kraus representation is not unique. The elements
of the Kraus operators can be written as vector components
\begin{equation}
m_j= \left(
\begin{array}{rc}
m_{1j}  \\
m_{2j}  \\
m_{3j}  \\
m_{4j}
\end{array}
\right)
\end{equation}
where j=0,1,2,3. The set of linear equations which need to be
satisfied is given in the Appendix. The stochastic map for the
squeezed vacuum channel(\ref{eq:image}) has Kraus operators which can be
realized in the following way:
\begin{eqnarray}
A_1 & = &  m_{10} I + m_{13} \sigma_3, \nonumber\\
A_2 & = &  (m_{21}+m_{22}) \sigma^\dagger + (m_{21}-m_{22}) \sigma,
\nonumber\\
A_3 & = &  m_{31} (\sigma^\dagger + \sigma),\\
A_4 & = &  m_{40} I \nonumber
\end{eqnarray}
with constants given by
\begin{eqnarray}
  m_{10}&=&\frac{1}{2}
       \frac{w_{eq}(1-\Lambda_3)}{\sqrt{1-\Lambda_1-\Lambda_2+\Lambda_3}},
  \nonumber \\
  m_{21}&=& \frac{1}{2}
  \frac{w_{eq}(1-\Lambda_3)}{\sqrt{1-\Lambda_1+\Lambda_2-\Lambda_3}},
  \nonumber \\
  m_{22}&=&-\frac{i}{2} \sqrt{1-\Lambda_1+\Lambda_2-\Lambda_3},
  \\
  m_{13}&=&\frac{1}{2} \sqrt{1-\Lambda_1-\Lambda_2+\Lambda_3},
  \nonumber \\
  m_{40}&=&\frac{1}{2}
  \sqrt{1+\Lambda_1+\Lambda_2+\Lambda_3-\frac{w_{eq}^2
       (1-\Lambda_3)^2}{1-\Lambda_1-\Lambda_2+\Lambda_3}},
  \nonumber  \\
  m_{31}&=&\frac{1}{2}
  \sqrt{1+\Lambda_1-\Lambda_2-\Lambda_3-\frac{w_{eq}^2
       (1-\Lambda_3)^2}{1-\Lambda_1+\Lambda_2-\Lambda_3}}.\nonumber
\end{eqnarray}
The Kraus decomposition is not unique so a different set of four
Kraus matrices can represent
the same map.  The minimum number of Kraus operators, for the map
constructed here, is four, but one could just as well use more
than four to produce the map.  This freedom in the Kraus
decomposition is related to the many possible ways of performing
measurements on the system.  The system is decohered as a result
of being measured by the environment.  The many different Kraus
operators correspond to the many different positive
operator-valued measures (POVMs) which lead to the decohered
state.  This is why there are many different sets of possible
Kraus operators which would result in the same map $\Phi$.

One can verify that this choice of Kraus operators satisfies the
condition in Eq. (\ref{eq:identity}) so that the final density
operator has trace unity.  This is a necessary condition for a
trace-preserving map, to which we restrict ourselves. However, one
could consider maps which are not trace-preserving in which case
Eq. (\ref{eq:identity}) becomes an inequality. Since the map is
non-unital, {\it i.e.}, it does not map identity into identity, we
have
\begin{equation}
  \Phi(I)=
  \left(
  \begin{array}{cc}
   1+w_{eq}(1-\Lambda_3)  &            0               \\
            0             &    1-w_{eq}(1-\Lambda_3)
  \end{array}
  \right).
\end{equation}
This is due to the presence of the affine shift.  Next, we discuss what
all this means geometrically in terms of the Bloch sphere.
\section{The Geometrical Picture}
For a set of general Bloch equations and a Lindbladian there exists a 4x4
matrix of the form (\ref{eq:tdmatrix}).  For the Lindblad equation we have
considered here,
this map can also be expressed, in terms of the 4x4 matrix, as
\begin{equation}    \label{eq:tdscmatrix}
  {\cal T_{\cal D}} =
  \left(
  \begin{array}{cccc}
    1                   & 0         & 0          & 0  \\
    0                   & \Lambda_1 & 0          & 0  \\
    0                   & 0         & \Lambda_2  & 0  \\
    (1-\Lambda_3)w_{eq} & 0         & 0          & \Lambda_3
  \end{array}
  \right).
\end{equation}
This is a special case of Eq. (\ref{eq:tdmatrix}) with
\begin{equation}
  0<\Lambda_3<\Lambda_1<\Lambda_2<1  \label{eq:lambdas}
\end{equation}
and affine shifts
\begin{equation}
  t_{01}=0,  t_{02}=0, t_{03}=w_{eq}(1-\Lambda_3),\hspace{.1in}
  -1<w_{eq}<0.
\end{equation}
From this we see that the components of the Bloch vector are contracted.
To guarantee that the Bloch vector is contained within the Bloch ball the
following condition must hold:
\begin{equation}
(\Lambda_1 u)^2 +(\Lambda_2 v)^2 + [\Lambda_3 w + (1-\Lambda_3)
w_{eq}]^2 \leq 1.
\end{equation}

Furthermore, we see that the matrix (\ref{eq:tdscmatrix})
transforms the Bloch sphere $u^2+v^2+w^2=1$ into an ellipsoid
inside the Bloch ball.  That is, the image of the set of pure
state density matrices under the stochastic map is given by the
family of ellipsoids
\begin{equation}
  \left( \frac{u}{\Lambda_1} \right) ^2 +
  \left( \frac{v}{\Lambda_2} \right) ^2 +
  \left(  \frac{w-w_{eq} (1-\Lambda_3)}{\Lambda_3} \right)^2=1.
\end{equation}
The shift, $w_{eq}(1-\Lambda_3),$ determines the center of the
ellipsoid while the eigenvalues,
$\lbrace\Lambda_1,\Lambda_2,\Lambda_3\rbrace,$ define the lengths
of the axes.  If we start with the set of pure state density
operators which lie on the Bloch sphere, then as time progresses,
the states move onto the surface of a contracting ellipsoid. The
pure states have become mixed states. By Eq. (\ref{eq:lambdas}),
each axis is unequally contracted, as shown in Figure 1.
Because
of the squeezed vacuum, the $v$-component experiences very little
damping while the $u$-component is rapidly damped, as shown in
Figure 2. The explicit expression for the shift in terms of
$w_{eq}$ indicates the ellipsoid is translated in the negative
$w$-direction over time and settles into an equilibrium or fixed
point, as shown in Figure 3.
\begin{widetext}
\begin{center}
\begin{figure}[H]
\centerline{\scalebox{0.7}
{\includegraphics{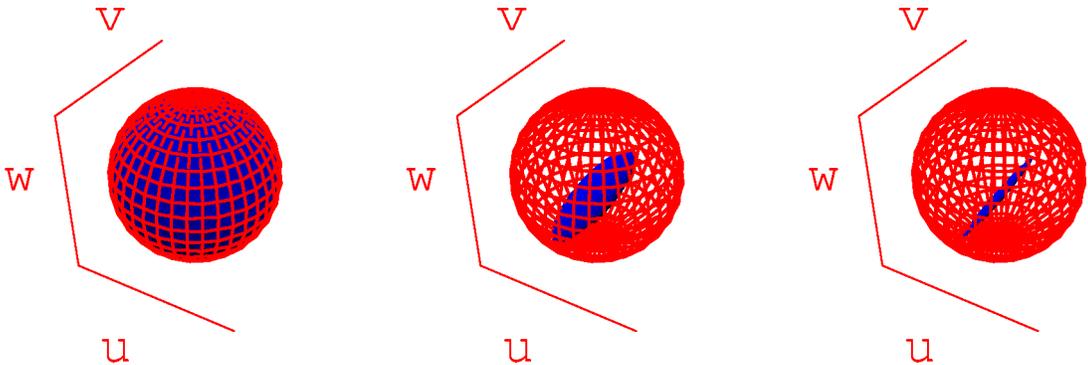}}}
\caption[short caption.]{The effect of noise on the set of qubit
density operators for an atom in a squeezed vacuum. Parameters are
N=1,\hspace{.1in} M=$\sqrt{2}$,\hspace{.1in} t=0,
\hspace{.1in} 0.5, \hspace{.1in} 1.}
\end{figure}
\begin{figure}[H]
\centerline{\scalebox{0.7}
{\includegraphics{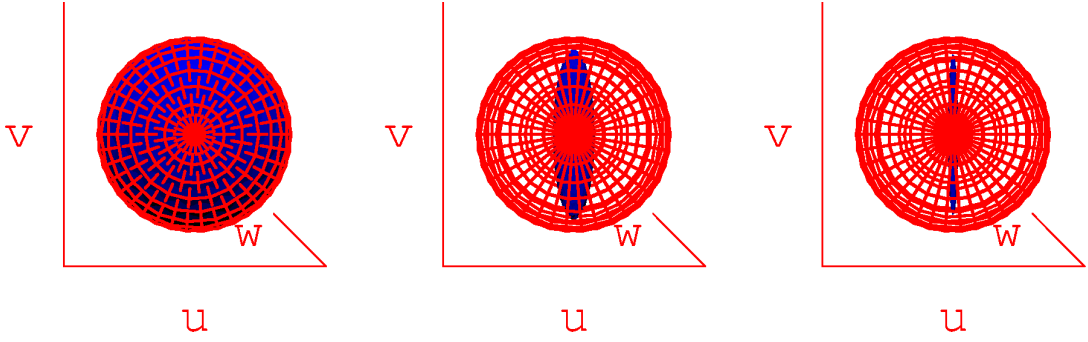}}}
\caption[short caption.]{From a top view one can see that the
u-component is rapidly damped while the v-component is slowly
damped. Parameters are N=1,\hspace{.1in}
M=$\sqrt{2}$,\hspace{.1in} t=0, \hspace{.1in} 0.5, \hspace{.1in}
1.}
\end{figure}
\begin{figure}[H]
\centerline{\scalebox{0.7}
{\includegraphics{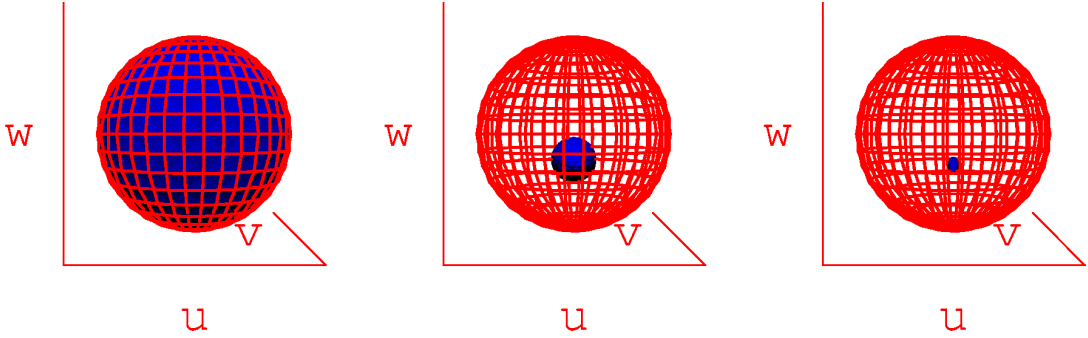}}}
\caption[short caption.]{A side view makes apparent the affine
shift of the non-unital map. Parameters are N=1,\hspace{.1in}
M=$\sqrt{2}$,\hspace{.1in} t=0, \hspace{.1in} 0.5, \hspace{.1in}
1.}
\end{figure}
\end{center}
\end{widetext}
The matrix of Eq.
(\ref{eq:tdmatrix}) is in the Pauli basis and describes the
dissipative dynamics from the stationary viewpoint at the center
of the Bloch sphere.  From this viewpoint, one would observe a
contracting ellipsoid moving away from the Bloch sphere center to
some equilibrium point.  The matrix could also be represented in
the damping basis, in which case it would be diagonal.  This
amounts to transforming to a frame where, instead of being a
stationary observer at the Bloch sphere center, one is moving
along with the shifting ellipsoid.  In this case, one observes
only a contracting ellipsoid.  This is the geometrical picture of
the damping basis.

If the effect of noise on all input states is known, then the
least noisy output states can be identified.  The minimal entropy
states are the states least affected by the noise and consequently
the least-mixed states. Geometrically, these would be all states
whose distance to the surface of the Bloch vector is a minimum.
These states form a set of extreme points of the convex set of density
operators.
Such a set of points on the ellipsoid having minimal distance to
the Bloch sphere may consist of one or
two points, a circle, or the entire surface of the ellipsoid.  For
a given noise, these nearest points represent states of maximum
purity in the set of states affected by the noise.  For the
squeezed vacuum channel, the set of minimal entropy states
consists of two states along the major axis of the ellipsoid.

The purest minimal entropy states are obtained by maximal
squeezing of the reservoir. This results in the most eccentric
ellipsoid which is stretched in one particular direction and
places the two points along the major axis of the ellipsoid close
to the Bloch sphere surface, as seen in Figure 1.  As the
squeezing parameter $M$ goes to zero, the ellipsoid becomes less
elongated and the minimal entropy states lose their purity.  When
M is identically zero, the minimal entropy points lie in a circle.
Equivalently, as M approaches zero, the damping for the $u$
component of the Bloch vector approaches that for the $v$
component.

If we had included the coherent dynamics as well as the
dissipative dynamics, the minimal entropy points on the ellipsoid
would be more mixed than without the coherent dynamics.  The
reason for this is that the ellipsoid would begin to rotate and
the minimal entropy points along the $v$-component would rotate
away from this axis where there is more noise.  Consequently, the
minimal entropy points get degraded by being moved away from the
axis with least noise so that for this channel, it is advantageous
to keep coherent dynamics suppressed.


Other noisy quantum channels have been introduced such as the
depolarizing channel, amplitude-damping channel, and the
phase-damping channel~\cite{nielsen2000}.  The depolarizing channel
has all three components of the Bloch sphere equally damped and it
is a unital channel.  The set of minimal entropy points would lie
on a sphere. The amplitude-damping channel has two of the three
components equally damped and it is a non-unital channel. It
arises from the master equation describing spontaneous emission as
in Eq. (\ref{eq:sponemission}). The set of all states moves on an
ellipsoid which shifts toward the South pole. The phase-damping
channel is the case described in Eq. (\ref{eq:phasedecay}). It has
two equally damped components, one undamped component, and it is
unital. It has two extreme points, one at the North pole and one
at the South pole.  Given any type of noise, the geometrical
picture is a useful aid and allows for a simple analysis of the
channel capacity.  To transmit information over the channel it is
ideal to use as input states those states which result in the
minimum amount of measurement error.

\section{Channel Capacity}

\subsection{Encoding Classical Information in Qubits}
One way to use a quantum channel is to send classical
information encoded in quantum states.  Although quantum
information is being sent because qubits are sent through the
channel, these qubits are being used in an entirely classical way.
A natural extension of classical channel capacity for a quantum
channel has been proposed by Holevo ~\cite{holevo1998}. In this
case, quantum states are used to transmit messages reliably across
some noisy quantum channel.  Also known as the product state
capacity, because messages are sent using tensor products of the
input states which comprise the alphabet, the channel capacity for
classical information over a noisy quantum channel has been
defined as
\begin{equation}
    {\mbox C(\Phi)}= \max_{(\textrm{p}_i,  \rho_i)}
    \lbrace \textrm{S}[\Phi(\sum_i {\mbox p}_i \rho_i)] - \sum_i
    \textrm{p}_i \textrm{S}[\Phi(\rho_i)] \rbrace.
    \label{eq:holevo}
\end{equation}
S is the von Neumann entropy defined as
$\textrm{S}(\rho)=-\textrm{Tr} (\rho\hspace{.03in}
\textrm{log}\hspace{.03in}\rho)$ and is the quantum analogue of
the Shannon entropy.  The maximum is taken over all possible
ensembles of input states.  The input alphabet consists of a set
of states $\rho_i$ which are transmitted with probability $p_i$.
For the case of the SVC, the
maximization is achieved using the two orthogonal input states for
which the damping is smallest with a uniform distribution over
these states. With the convention already chosen, this is the
v-component of the Bloch sphere.
Thus, the optimal way to send classical information through the SVC is
to prepare product states using the two orthogonal input states $\rho_0=|0
\rangle \langle 0 |_v$ and $\rho_1=|1 \rangle \langle 1 |_v$,which lie in the
equatorial plane of the Bloch sphere, with
corresponding probabilities ${\mbox p}_0 = 1/2$ and ${\mbox p}_1 =
1/2$ to encode messages.  For example,
$\rho_0^{(1)}\otimes\rho_1^{(2)}\otimes \cdots\otimes\rho_1^{(N)}$,
requiring N uses of the channel, would represent the N-length
classical bit string 01$\cdots$1.

The
noise operation $\Phi$ causes the output of these two states to be
non-orthogonal, mixed states. No single measurement at the
receiving end can perfectly determine which input state was sent.
There are two errors which can occur, one error arises because of
the non-distinguishability of non-orthogonal states while the
other arises because the state is mixed. The first term in Eq.
(\ref{eq:holevo}) takes into account the information loss due to
the fact the the output states are non-orthogonal while the second
term takes into account the fact the the output states are mixed.
The Holevo capacity has a nice geometrical interpretation. To see
this we consider an example.  For simplicity, let
us assume that the SVC has maximum squeezing and the input states
are received after a one second pass through the channel.  Then,
using the same parameters as in Figures 2-4, namely,
N=1,M=$\sqrt{2}$, and t=1, we find explicitly that the Holevo
capacity is C(SVC)=.93-.11=.82 qubits per
transmission.  This is calculated by using $\rho_0=|0
\rangle \langle 0 |_v$ and $\rho_1=|1 \rangle \langle 1 |_v$ as input states
with corresponding probabilities ${\mbox p}_0 = 1/2$ and ${\mbox p}_1 =
1/2$ and calculating the quantity inside the brackets of Eq. (\ref{eq:holevo}).

Note that the first term is present because of the
affine shift while the second term is due to a contraction of the
Bloch sphere.  To see the various errors which can occur we can
write C(SVC)=1-(1-.93)-.11=.82 qubits per transmission.  Now the first
term is the number of qubits which can be sent if there was no
noise present.  The second term is the error caused by the affine
shift. This is a distance from the center of the Bloch sphere to
the center of the ellipsoid. This error occurs because the states
lose their orthogonality. The last term is the error due to the
contraction of the ellipsoid. The contractions cause the states to
become mixed and consequently there will be measurement errors of
this kind.  This term is a distance from the Bloch sphere surface
to the surface of the ellipsoid. It is interesting to note that
the affine shift actually takes the maximally mixed state into a
state with less entropy.  This can occur because generalized
measurements can decrease entropy. Also, if no noise were present
the capacity would be C(\textsc{I})=1 qubit per transmission
meaning that one error-free qubit can be transmitted in one use of
the channel.

\subsection{Entanglement Transmission}
Another way in which a quantum channel can be used which does not
have a classical counterpart is for entanglement transmission.  In
this case, one is interested in distributing parts of an entangled
state to different locations.  For example, a source may generate
EPR pairs and one may be interested in sending one half of this
EPR pair through some channel to a receiver.  Naturally, noise
will corrupt the transmitted state and lead to a decrease in the
entanglement of the joint state.

A channel has the capacity to transmit entanglement if after
passing through the channel there is any nonzero entanglement
present in the joint state.  It has been shown by Horodecki
\textit{et. al} that any nonseparable bipartite system which has
entanglement, however small, can be distilled to a singlet form
~\cite{horodecki1997}. Therefore, if a channel can transmit any
entanglement, it is a useful channel.  N copies can be sent and a
singlet state can be distilled.

Assuming the initial Bell state has the following density operator
representation
\begin{equation}
\rho_{AB}=\frac{1}{2} \left(
\begin{array}{cccc}
1 & 0 & 0 & 1 \\
0 & 0 & 0 & 0 \\
0 & 0 & 0 & 0 \\
1 & 0 & 0 & 1
\end{array}
\right)
\end{equation}
The output state will be determined by the following operation
\begin{equation}
    \rho'_{AB}= \mathbf{1}_A \otimes \Phi_B [ \rho_{AB} ]
\end{equation}
which describes the process of sending one qubit to Bob while
Alice keeps her qubit intact.  In matrix notation, the output
state for the joint system is
\begin{widetext}
    \begin{equation}
    \rho'_{AB}=\frac{1}{4} \left(
    \begin{array}{cccc}
    1+w_{eq}+\Lambda_3 (1-w_{eq}) & 0 & 0 & \Lambda_1+\Lambda_2 \\
    0 & 1-w_{eq}-\Lambda_3 (1-w_{eq}) & \Lambda_1-\Lambda_2 & 0 \\
    0 & \Lambda_1-\Lambda_2 & 1+w_{eq}-\Lambda_3 (1+w_{eq}) & 0 \\
    \Lambda_1+\Lambda_2 & 0 & 0 & 1-w_{eq}+\Lambda_3 (1+w_{eq})
    \end{array}
    \right).
    \end{equation}
\end{widetext}
At the initial time the joint state is maximally entangled and the
noise process should result in a decrease in the entanglement
until some critical time when the joint state is separable and
remains separable thereafter.  Using the Peres criterion of the
positivity of the partial transpose, one can determine when
the state becomes separable~\cite{peres1996}. A necessary and sufficient condition
for the output state to be nonseparable is that the partial
transpose map be negative~\cite{horodecki1996}. To check the
positivity of the partial transpose it suffices to examine the
eigenvalues of the operator given by
\begin{equation}
    \rho'^{T_B}_{AB}=\mathbf{1}_A \otimes \mathrm{T}_B
    [\rho'_{AB}]
\end{equation}
where $\mathrm{T}_B$ denotes the transpose of the state of Bob's
qubit. The four eigenvalues of the partial transpose matrix are
\begin{eqnarray}
    e_1 &=& \frac{1}{4} \{ 1+\Lambda_3-\sqrt{(\Lambda_1-\Lambda_2)^2+[w_{eq}(1-\Lambda_3)]^2}
    \} \nonumber \\
    e_2 &=& \frac{1}{4} \{ 1+\Lambda_3+\sqrt{(\Lambda_1-\Lambda_2)^2+[w_{eq}(1-\Lambda_3)]^2} \}
    \\
    e_3 &=& \frac{1}{4} \{ 1-\Lambda_3-\sqrt{(\Lambda_1+\Lambda_2)^2+[w_{eq}(1-\Lambda_3)]^2} \}
    \nonumber \\
    e_4 &=& \frac{1}{4} \{ 1-\Lambda_3+\sqrt{(\Lambda_1+\Lambda_2)^2+[w_{eq}(1-\Lambda_3)]^2} \}
    \nonumber.
\end{eqnarray}
Initially, the eigenvalues have values $e_1=1/2, e_2=1/2,
e_3=-1/2,$ and $e_4=1/2$ while in steady state they become
$e_1=1/6, e_2=2/6, e_3=1/6,$ and $e_4=2/6$. We find that the
nonseparability is determined solely by the eigenvalue $e_3$. The transmitted EPR
state remains nonseparable provided
\begin{equation}
    1-\Lambda_3 <
    \sqrt{(\Lambda_1+\Lambda_2)^2+[w_{eq}(1-\Lambda_3)]^2}.
\end{equation}
There is some critical time when the eigenvalue is zero and then
remains positive thereafter.  The value of this critical time
depends on the parameters of the reservoir.  As the photon number
of the reservoir increases
the critical time decreases.  One would expect this to be the case since the reservoir is more
noisy.  It is not as obvious what the effect of the squeezing
of the reservoir has on the entanglement because the squeezing results in a
tradeoff of more noise in one component and a decrease in another component.
We find that a squeezed reservoir results in a longer entanglement time for the
inital maximally entangled state.  The critical time is longer for a maximally squeezed vacuum.
This is shown by the solid curve in Fig. 4.
\begin{figure}[htbp]
\centerline{\scalebox{0.7}{\includegraphics{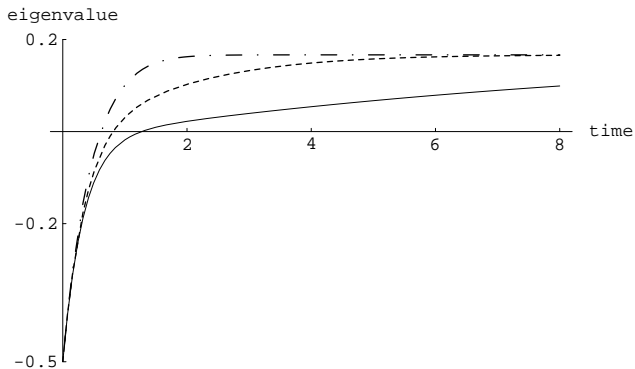}}}
\caption[short caption.]{The eigenvalue $e_3$ is shown as a
function of time for photon number N$=1$.  The three curves
correspond to squeezing parameter $M=0$(dash-dotted), $M=.8
M_{max}$(dashed), and $M=M_{max}.$ The larger the squeezing
parameter, the longer the joint state is nonseparable.}
\end{figure}
In both the case of sending product states and for distributing EPR states,
the squeezing parameter, M, is able to enhance the capacity of the channel.

\section{Conclusion}
In this paper we provided a method for calculating
a stochastic map for any quantum Markov channel.  Our method
uses a special damping basis of left and right eigenoperators.
This basis provides a natural way to generate noisy quantum channels
from a quantum optical approach.  This technique allows one to
calculate the stochastic map and from this one can then determine a
set of Kraus operators which define the noisy quantum channel.

We used this method to calculate explicitly a noisy quantum channel
we called the squeezed vacuum channel.  We showed the relationship between
a set of quantum optical Bloch equations and the damping basis.
Some known quantum channels such as the amplitude-damping channel and
the depolarizing channel arise from this set of Bloch equations.  These
quantum Markov channels are special cases of the squeezed vacuum channel.
The channel we derived is a non-unital stochastic map which is characterized
by three unequal damping eigenvalues.  By using the damping basis, we were
able to find a set of Kraus operators for the stochastic map.  From this,
the effect of noise on the set of input states was
interpreted geometrically.  The Bloch picture was used to study
the effect of noise present in the channel and the
coherent dynamics was considered in addition to the incoherent dynamics.

The procedure to calculate the channel capacity requires a
maximization over all input states.  Using the Bloch picture, we
were able to determine the channel capacity for the
squeezed vacuum channel to transmit classical information in
quantum states. We found that the channel has two minimal entropy
points which should be used to optimally transmit information.  A
geometrical interpretation of the Holevo capacity was given and
used to identify two types of errors--those arising from
non-orthogonal states and those arising from mixed states. We also
discussed the ability of this channel to distribute EPR states. We
found that, after sending one half of the EPR state through the
channel, there is some critical time after which the state becomes
separable.

This paper shows that with the {\it a priori} knowledge of the
effect of noise given by a Lindblad form, one can choose to encode
messages using the pure states which are closest to the final
states of minimum entropy.  The squeezed vacuum channel is a more
general noisy channel derived from quantum optical two-level
systems. Unlike previously introduced channels, it has unequal
damping eigenvalues and it is non-unital.  We found that channel
capacity is enhanced by the squeezing parameter, M, whether it is used to send
product states or used to transmit EPR states. This channel may prove
useful as a testing ground for future conjectures on quantum
channel capacities.
\begin{acknowledgments}
We thank Dr. G.H. Herling for
discussions and corrections to the final draft.  This work was
partially supported by a KBN grant No. 2PO3B 02123
and the European Commission through the Research Training Network
QUEST.\\
\end{acknowledgments}
\section{Appendix}
For the Hilbert space of 2x2 matrices, one may always represent
the quantum channel using four or less Kraus operators. To find a
representation, one can construct a set of simultaneous equations
via the prescription given in Section VI.B. The squeezed vacuum
channel has Kraus operators which must satisfy the following set
of equations. There are four for the shifts:
\begin{eqnarray}   \label{eq:linearset}
m_0^\star \cdot m_0 + m_1^\star \cdot m_1 + m_2^\star \cdot m_2
     +m_3^\star \cdot m_3  &=& 1  \\  \nonumber
m_0^\star \cdot m_1 +m_0 \cdot m_1 ^\star + i (m_2^\star \cdot m_3
     - m_2 \cdot m_3^\star ) &=& t_{01} \\   \nonumber
m_0^\star \cdot m_2 +m_0 \cdot m_2 ^\star - i (m_1^\star \cdot m_3
     - m_1 \cdot m_3^\star ) &=& t_{02} \\   \nonumber
m_0^\star \cdot m_3 +m_0 \cdot m_3 ^\star + i (m_1^\star \cdot m_2
     - m_1 \cdot m_2^\star ) &=& t_{03}
\end{eqnarray}
For the coefficients of $a,c,d,$ and $\star{d}:$
\begin{widetext}
\begin{eqnarray}
i (m_0^\star \cdot m_2 - m_0 \cdot m_2^\star ) - (m_1^\star \cdot
m_3 + m_1 \cdot m_3^\star ) &=& 0 \nonumber  \\   \nonumber
m_0^\star \cdot m_0 +m_1^\star \cdot m_1 -m_2^\star \cdot m_2 -
m_3^\star \cdot m_3 + i (m_1^\star \cdot m_2 +m_1 \cdot m_2^\star
) -
(m_0^\star \cdot m_3 - m_0 \cdot m_3^\star) &=& \Lambda_1   \\
\nonumber m_0^\star \cdot m_0 + m_1^\star \cdot m_1 -m_2^\star
\cdot m_2 - m_3^\star \cdot m_3 -i( m_1^\star \cdot m_2 +m_1 \cdot
m_2^\star )
+ (m_0^\star \cdot m_3 - m_0 \cdot m_3^\star ) &=& \Lambda_1  \\
\nonumber m_0^\star \cdot m_1 - m_0 \cdot m_1^\star - i (m_2^\star
\cdot m_3 + m_2 \cdot m_3^\star ) &=& 0  \nonumber  \\
m_0^\star \cdot m_0 - m_1^\star \cdot m_1 +m_2^\star \cdot m_2
-m_3^\star \cdot m_3 - i (m_1^\star \cdot m_2 +m_1 \cdot
m_2^\star) - (m_0^\star \cdot m_3 - m_0 \cdot m_3^\star ) &=&
\Lambda_2   \\  \nonumber m_0^\star \cdot m_0 - m_1^\star \cdot
m_1 +m_2^\star \cdot m_2 -m_3^\star \cdot m_3 + i (m_1^\star \cdot
m_2 +m_1 \cdot m_2^\star) + (m_0^\star \cdot m_3 - m_0 \cdot
m_3^\star ) &=& \Lambda_2  \\  \nonumber m_0^\star \cdot m_0 -
m_1^\star \cdot m_1 - m_2^\star \cdot m_2 +m_3^\star \cdot m_3 &=&
\Lambda_3  \\  \nonumber m_0^\star \cdot m_1 - m_0 \cdot m_1^\star
+m_1^\star \cdot m_3 +m_1 \cdot m_3^\star +i ( m_0^\star \cdot m_2
- m_0 \cdot m_2^\star +m_2^\star \cdot m_3 +m_2 \cdot m_3^\star )
&=& 0  \\  \nonumber m_0^\star \cdot m_1 - m_0 \cdot m_1^\star
-m_1^\star \cdot m_3 -m_1 \cdot m_3^\star +i (- m_0^\star \cdot
m_2 + m_0 \cdot m_2^\star +m_2^\star \cdot m_3 +m_2 \cdot
m_3^\star ) &=& 0  \\  \nonumber
\end{eqnarray}
\end{widetext}
And to satisfy the condition $A_k A_k^\dagger = I$:
\begin{equation}
m_0^\star \cdot m_3 +m_0 \cdot m_3^\star -i (m_1^\star \cdot m_2 -
m_1 \cdot m_2^\star) = 0
\end{equation}
The set of equations leads to restrictions on the damping
eigenvalues as in Eq.(\ref{eq:inequalities}).  As an example of
the explicit construction of Kraus matrices, we will consider the
pure phase decay case of Eq.(\ref{eq:phasedecay}) which is the
phase-damping channel. For this case, $\Lambda_1=\Lambda_2$ and
$\Lambda_3=0$ and $t_{01}=t_{02}=t_{03}=0$.  Any solution of the
set of linear equations along with these conditions will give a
Kraus representation.  The following solution for elements of the
Kraus matrices is easily found to be $m_2=\vec{0}$, $m_3=\vec{0}$
and
\begin{equation}
m_0= \left(
\begin{array}{c}
m_{10}  \\
0  \\
0  \\
0
\end{array}
\right) , \hspace{.1in} m_3= \left(
\begin{array}{c}
0  \\
0  \\
0  \\
m_{43}
\end{array}
\right)
\end{equation}
with
\begin{equation}
  m_{10}=\sqrt{\frac{1+\Lambda}{2}}, \hspace{.2in}
  m_{43}=\sqrt{\frac{1-\Lambda}{2}}.
\end{equation}
The phase-damping channel has a Kraus representation given by:
\begin{eqnarray}
  A_1 &=& \sqrt{\frac{1+\Lambda}{2}} \hspace{.1in} {\mbox I} \\
  A_2 &=& \sqrt{\frac{1-\Lambda}{2}} \hspace{.1in} \sigma_3
\end{eqnarray}


\end{document}